\documentclass[11pt]{article}
\usepackage{epsfig}

\begin{document}
\newcommand{\ov}{\overline}

\title{At the boundary between biological and cultural
evolution: The origin of surname distributions}

\author{Susanna C. Manrubia \\ {\it Centro de Astrobiolog\'{\i}a,
INTA-CSIC} \\{\it Ctra. de Ajalvir km. 4, 28850
Torrej\'on de Ardoz, Madrid, Spain} \\ \\
Dami\'{a}n H. Zanette \\ {\it Consejo Nacional de Investigaciones
Cient\'{\i}ficas y T\'{e}cnicas} \\ {\it Centro At\'{o}mico
Bariloche and Instituto Balseiro, 8400 Bariloche, Argentina}}

\date{\today}

\maketitle

\begin{abstract}

Surnames and nonrecombining alleles are inherited from a single
parent in a highly similar way. A simple birth-death model with
mutations can accurately describe this process. Exponentially
growing and constant populations are investigated, and we study
how different compositions of the founder populations can be
observed in present-day diversity distributions. We analyse
different quantities in the statistically stationary state, both
through analytic and numerical methods. Our results compare
favourably to field data for family sizes in several countries.
We discuss the relationship between the distribution of surnames
and the genetic diversity of a population.

\end{abstract}

\baselineskip .2in

\section{Introduction}

Biological  and  cultural  features of human populations have been
traditionally   studied   by   separate   disciplines,   but   the
parallelisms  between  biological and cultural evolution have been
put  forward  by  a  number  of researchers. Already Darwin (1871)
pointed  out  that {\it ``the formation of different languages and
of  distinct species and the proofs that both have been developed
through a gradual process are curiously parallel.''}

Cultural   traits   are   transmitted   from  ancestors  to  their
descendence,  in  a  process  analogous  to  inheritance,  and are
subject  to  changes, similar to mutations, by interaction between
individuals  --such  as  teaching  and  imitation.  Moreover, they
usually  fulfill  a  practical  purpose,  which  amounts  to being
subject  to  selection.  In  fact,  they enhance the relationships
within   human  groups,  defining  social entities  comparable  to
certain  biological species and populations. The quantification of
cultural  traits  has  been  attempted only recently. For
example,  Cavalli-Sforza {\it et al.} (1982) applied concepts from
the  quantitative  theory  of  biological evolution to construct a
theory  of cultural evolution. They analyzed forty traits, ranging
from  political preferences to superstitions. Many of these traits
are  subject  to  high  mutability,  since  they are influenced by
fashion,  individual  acquaintances,  and personal experience, and
one  does  not expect their quantitative properties to be directly
comparable  to  those  of any biological feature. Other traits, on
the other hand, are better preserved. Among them we find languages
and  surnames.  Language  is essential to integrate the individual
to  society;  surnames  are  historical --though recent-- signs of
identity  in  social groups. Quite early, Galton and Watson (1874)
dealt  with  the  problem  of  the  extinction  of  surnames. This
problem  is  equivalent  to  that  of  the  extinction of a mutant
allele   in   a   population,  although  this  relation  was  only
established  half  a  century later (Fisher, 1922; Haldane, 1927),
when   the   first   quantitative  approaches  to biological
evolutionary processes took place.

Comparative  methods  analogous to biological taxonomy are used to
determine  the degree of similarity between languages. This method
returns  the  {\it genetic classification} of linguistic diversity
(see  for  example  Greenberg,  1992;  Ruhlen,  1992). Recently, a
quantitative  study  of the taxonomy of languages has been carried
out   (Zanette,  2001),  showing  that  the  distribution  of  the
number  of  subtaxa  within  a  taxa  displays scaling properties,
quantitatively   similar   to   those   disclosed   in  biological
taxonomy  by  Burlando  (1990,  1993).  That  is,  if  $n$  is the
number  of  subtaxa  belonging  to  a given taxa --say, the number
of  languages  in  the  Indo-European  family,  or  the  number of
species  in  the  genus  {\it Canis}-- the fraction $p(n)$ of taxa
that   have   precisely   $n$   subtaxa   scales   as  $p(n)  \sim
n^{-\beta}$.    The    exponent   is   found   in   the   interval
$1\leq\beta  \leq  2$  in  both  cases.  This  gives  quantitative
support  to  Darwin's  observation  on the ``equivalence'' between
the  mechanisms  behind  biological  and  linguistic  evolution. A
complementary   comparison   is   that  of  linguistic  abundance,
measured  as  the  number  of  individuals  speaking  a  language,
with  the  number  of  individuals of a biological species. Again,
the  frequency  as  a  function  of  the number of individuals has
scaling  properties  both  for  languages  (Gomes  {\it  et  al.},
1999) and for species (Pielou, 1969), see Fig. 1.

Surnames  are  cultural  traits  (Cavalli-Sforza \& Feldman, 1981)
whose  transmission  bears  strong  similarity  with  that of some
biological   features.   They   are  inherited  from  one  of  the
parents,  usually  the  father,  much  in  the  same  way  as  the
Y-chromosome   or   the   mitochondrial  DNA.  The  extinction  of
a  surname  and  the  persistence  of  a  non-recombining  neutral
allele   are   equivalent   problems.   This   is   not   only   a
mathematical  fact,  but  has also practical implications. Indeed,
to  assess  the  multiple  or  single  origin of a surname one can
turn  to  genetic  measures,  since males sharing the same surname
might   also  share  the  same  haplotype  in  the  nonrecombining
segment  of  the  Y  chromosome  (Sykes \& Irven, 2000). In a very
large  population,  the  statistical  properties  of  the  surname
distribution    can    be   strongly   correlated   with   genetic
diversity  (Barrai  {\it  et  al.}, 1996), and may even be used to
understand  the  genetic structure of a population (Yasuda {\it et
al.},  1974).  Recent  reports  on  actual  populations  (Miyazima
{\it  et  al.},  2000;  Zanette  \&  Manrubia, 2001) show that the
distribution   of   surnames  follows  the  same  statistical  law
observed  for  languages and biological species. Namely, if $n$ is
the   number   of   individuals   bearing  a  given  surname,  the
fraction  $p(n)$  of  surnames  decreases  with  $n$ as $p(n) \sim
n^{-\gamma}$  (see  Fig. 1).  Here,  however, the exponent is
always   $\gamma   \approx   2$.  This  paper focuses  on  a
model  aimed  at predicting this kind of regularities, observed
in disparate human populations.

\begin{figure} \label{fig:f1}
\centerline{\psfig{file=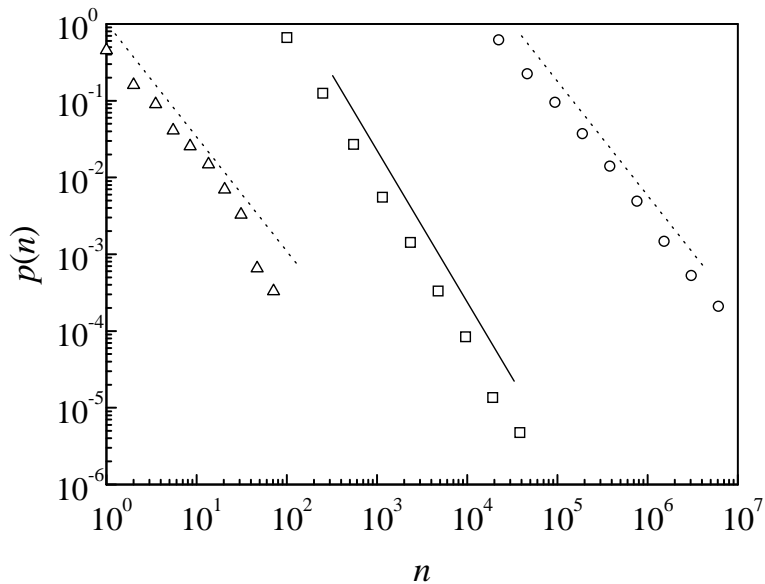,width=12 cm}}
\caption{Scaling behaviour of the fraction $p(n)$ of species
represented by $n$ individuals (triangles), of surnames borne by
$n$ persons (squares), and of languages with $n$ speakers
(circles). Data for the distribution of species abundance are
from Poore (1968), corresponding to trees in a Malaysian
rainforest [see also Sol\'e and Alonso (1998)]. Data for surnames
(beginning by A) are from the 1996 Berlin telephone book. Data
for languages are from {\tt
http://www.sil.org/ethnologue/preface.html}. As a guide to the
eye we draw lines with slope $-2$ (solid line) and $-3/2$ (dotted
lines). Data sets have been mutually shifted for better
visualization.}
\end{figure}

\section{Theoretical approaches to surname evolution}

At  the  mathematical level, several models have been proposed and
analysed  in order to identify quantitative properties of surnames
evolution  --or,  equivalently,  those  of  nonrecombining neutral
alleles.  The  main  points addressed by these studies are (i) the
probability  of  fixation  of  a  given surname/allele in a closed
population  (Galton  \& Watson, 1874; Fisher, 1922; Haldane, 1927;
Moran,  1962;  Lange,  1981;  Rannala  1997;  Hull 1998), and (ii)
the  distribution  of  the  number of individuals bearing the same
surname/allele  (Kimura  \&  Crow, 1964; Karlin \& McGregor, 1967;
Fox \& Lasker, 1983; Panaretos, 1989; Gale 1990; Consul, 1991;
Islam,
1995;  Zanette  \&  Manrubia,  2001).  Indeed, these two questions
cover   complementary   aspects  of  the  same  problem.  In  (i),
one  deals  with  a  closed  population  (no  immigrants enter the
system),  and  implicitly  assumes  that  the mutation rate is low
enough,  such  that fixation can indeed occur before a mutant form
appears.   Suppose   that   there   are  $N$  individuals  in  the
population.  We  know  from coalescence theory that the time $n_g$
(in  units  of  the  number of generations) required for a neutral
allele  to be fixed is of order $n_g \sim N$. Now suppose that the
mutation  rate  per  generation and per individual is $r$. Then $M
\equiv  r  n_g  N$  is  the  average number of mutants after $n_g$
generations  have  elapsed.  Only  if $M \ll 1$, that is if $r \ll
N^{-2}$,  will  the fixation of the allele be possible. As soon as
this  inequality  is  violated,  a  new situation arises, in which
both  neutral  drift  and  mutation  play  relevant roles. In this
case,  a  broad  distribution  of surnames or of genetic diversity
are  expected. Actually, the value of $r$ will be usually fixed by
the  nature  of  the  problem, while the size $N$ of the considered
population  can  increase enough such that $N^2 > 1/r$. Therefore,
the  statistical  behaviour  crosses  over  to  the second regime,
where  the  appearance of mutants cannot be discarded. We are then
in the assumptions of the models of class (ii).

All  the  models  in  class  (ii)   represent   systems  where
statistical  equilibrium  is  reached.  While  the  neutral  drift
drives  the  less  frequent  surnames out of the system, mutations
generate  new  surnames.  For large times, the number of different
surnames  in  the  system  is  much  larger than unity. Yasuda and
co-workers  (Yasuda  {\it  et  al.}, 1974) used a stochastic model
for  a  population  with  a  fixed  number of individuals. At each
evolution   step,  a  new  individual  with  the  surname  of  his
father  is  added,  but  he  acquires  a  new surname with a given
probability.   In  any  case,  a  randomly  chosen  individual  is
eliminated  in  order  to  keep  the  population  constant.  Their
analytical  results  compared  successfully  to field data despite
the   restriction   of   constant  population,  which  forces  the
elimination  of  individuals.  In  this  model,  the  size  of the
progeny is Poisson distributed.

Other  models start with a single individual in the population and
new  individuals  carrying  the  same  surname  or  a  new one are
sequentially  added  (Panaretos, 1989; Consul, 1991; Islam, 1995).
These  fall  in the category of branching processes (Harris, 1963)
with  an  increasing  number  of  individuals  in  the population.
Panaretos  (1989)  rephrased a model introduced by Simon (1955) in
the   context  of  linguistics,  as  follows.  At  each  step,  an
individual  is  chosen  to  be  the  father  of  a newborn and his
surname  is  transmitted  with  probability  $1-\alpha$.  With the
complementary  probability  $\alpha$,  the  newborn  is assigned a
surname  not present in the population. Recently, we have modified
Simon's  model  by  (i)  introducing an additional parameter $\mu$
which  represents  the  death  rate  of  the  individuals  in  the
population,  and  (ii)  allowing for arbitrary initial conditions.
Preliminary  results  have  been  successfully compared with large
data  sets (Zanette \& Manrubia, 2001). This is the starting point
of   the  present  work, where we present new analytical and
numerical results for this birth-death model.

We  describe  the  model in the next section, where our analytical
results  are  derived.  Questions  like  the  role  played  by the
composition  of  the  founder population and by death events in an
exponentially  growing population are addressed. We also analyse a
system    with   constant   population,   which   turns   out   to
quantitatively differ from the previous case. In Section \ref{num}
we  numerically  check  our  analytical  results  and run computer
simulations  of  the  model  in  situations  not  amenable  to  an
analytical description. In Section \ref{empirical} our results are
tested  against  field  data. We finish with an overall discussion
in the last section.

\section{Birth-death model for surname evolution}
\label{model}

Our  model  population  evolves  in  discrete  steps,  each   step
corresponding to  the birth  of a  new individual.   At each  time
step, moreover, an individual is  chosen at random from the  whole
population   and   becomes   removed   with   probability   $\mu$,
representing a  death event.   The total  population at  step $s$,
$P(s)$,  is  therefore  a  stochastic  process  governed  by   the
evolution equation
\begin{equation} \label{Pest}
P(s+1)=P(s) +1-w(s),
\end{equation}
where $w(s)$ is the dichotomic stochastic process
\begin{equation} \label{xi}
w (s)=\left\{
\begin{array}{ll}
1 & \mbox{with probability $ \mu$}, \\
0  & \mbox{with probability $1-\mu$}.
\end{array}
\right.
\end{equation}
We show in Appendix \ref{ap1} that, averaging over realizations of
the stochastic process $w(s)$,  the average total population  $\ov
P$ grows exponentially in time:
\begin{equation} \label{Pt}
\ov P(t) = P_0 \exp[\nu(1-\mu) t].
\end{equation}
Here, $\nu$ is  the birth rate  per individual and  unit time, and
the product $\nu  \mu$ turns out to  be the corresponding  death
rate; $P_0$ is the initial population.

We   think   of  the  population  as  divided  into  groups  --the
families--  within  which  all  individuals bear the same surname.
At  each  birth  event, the newborn is assigned a new surname, not
previously   present   in   the   population,   with   probability
$\alpha$.  This  probability  can  be  seen as a mutation rate for
surnames,  but could also be interpreted as a measure of migration
towards  the population of individuals with new surnames. With the
complementary  probability,  $1-\alpha$,  a preexistent individual
is   chosen   at  random  from  the  population  and  becomes  the
newborn's  father,  i.e.  his  surname  is  given  to the newborn.
Surnames     are    therefore    assigned    with    probabilities
proportional   to   the   size   of  the  corresponding  families,
allowing   however   for   fluctuations  inherent  to  the  random
distribution   of   births   among   families.  Consequently,  the
distribution    of    surnames   is   driven   by   a   stochastic
multiplicative  process  (Van  Kampen,  1981),  modulated  in turn
by  the  total  population  growth.  This  process is analogous to
the  mechanism  proposed  by  Simon  (1955) to model the frequency
distribution  of  words  and  city  sizes, described by Zipf's law
(Zipf,  1949),  among other instances. Our model, in fact, reduces
to  Simon's  model  if  mortality  is neglected, i.e. for $\mu=0$.
Note  that  allowing  for  death  events  adds  a relevant process
in  the  case  of surnames, namely, the possibility that a surname
disappears  if  it  is  borne  by  a  single  individual  and  the
individual dies.

Since  neither the probability of becoming father of a newborn nor
the  death probability depend on the individual's age, the present
model  population  can be thought of as ageless. As a consequence,
if   $\nu$  and  $\mu$  are  constant,  the  probability  that  an
individual dies at an age between $T$ and $T+dT$ is
\begin{equation}    \label{pT}
dp(T) = \nu\mu \exp(-\nu\mu T) dT,
\end{equation}
which implies a life expectancy $\ov T = 1/\nu\mu$. Moreover,  the
probability that an individual has exactly $m$ children with the
same family name during its whole life equals
\begin{equation}    \label{pm}
p(m)=  (1-\alpha) \mu [1+(1-\alpha) \mu]^{-m-1},
\end{equation}
giving a fertility $\ov m = 1/(1-\alpha) \mu$. The exponential
distribution in  (\ref{pm})  is  in  reasonable  agreement  with
actual   data collected over  relatively short  periods, for
instance, for  the United  States  (Hull,  1998),  but  contrasts
with the Poissonian distribution found in data  integrated over
several centuries,  as in  the  case  of  England  (Dewdney,
1986). This discrepancy can presumably be ascribed to long-range
variations of real birth  and death rates (see Fig. 2).

We are assuming that an individual's surname is inherited from the
father.  Consequently,  the model --as presented above-- describes
the  evolution  of  the  male  population only. The same evolution
rules   apply   however   if  it  is  assumed  that  surnames  are
transmitted  with  the  same probability by either parent. In this
case,  there  is  no sex distinction and the model encompasses the
whole  population.  The  real  situation  is  in fact intermediate
between  these two limiting cases: whereas some societies strongly
favor  inheritance  of  the surname either from the father or from
the  mother,  other  groups  allow  for  a  choice between the two
possibilities.  In  some  developed  western  countries, where the
surname  is  preferentially  transmitted by the father, opting for
the  mother's  surname  has been proposed as a method to avoid the
persistent  extinction  of  surnames,  due  to  the extremely slow
population growth (Legay \& Vernay, 1999).

\begin{figure}\label{fig:f2}
\centerline{\psfig{file=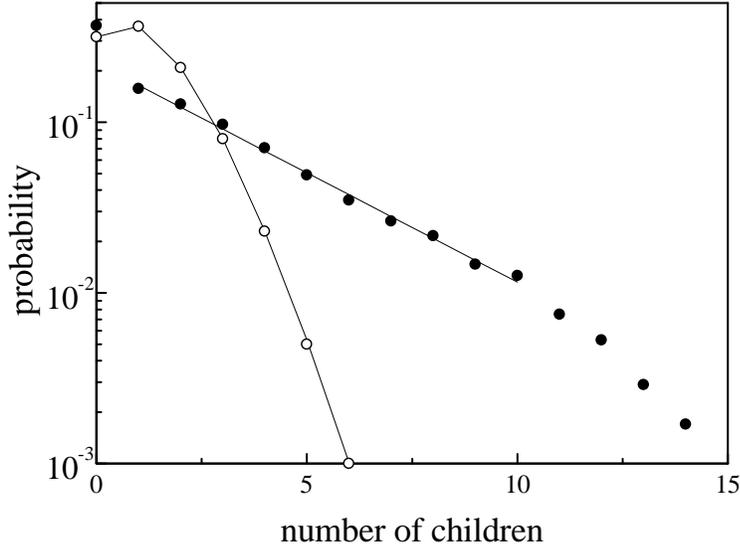,width=12 cm}}
\caption{Probability distribution  of the  number of  children per
male  in  two  different  populations.  Open dots correspond to an
average during  the period  1350-1986 in  England (Dewdney, 1986).
The solid line joining the data points is actually a fitting  with
a  Poisson  distribution,  $P(n)=\exp(-\lambda) \lambda^n/n!$ with
average $\langle n\rangle=\lambda=1.15$. Solid dots are data  from
the 1920 American census [from Hull (1998); originally compiled by
Lotka (1931)]. The solid line  fitting the first part of  the data
is an exponential distribution, $P(n)\propto \exp(-0.3 n)$.}
\end{figure}

\subsection{Distribution of families by size. The case of $\mu=0$}

As stated above, our model reduces to Simon's model (Simon,  1955)
for $\mu =0$.   In this case, $w(s)=0$  for all $s$ and  the total
population evolves deterministically, $P(s)=P_0+s$, since  exactly
one individual is added to the population at each step. The number
of families with  exactly $i$ individuals  at step $s$,  $n_i(s)$,
grows to  $n_i(s)+1$ when  an individual  is added  to a family of
size $i-1$.  This occurs with probability $(1-\alpha)(i-1) /P(s)$.
On the average, thus,  the number of families  with$i$ individuals
varies according to
\begin{equation}   \label{Simon2}
\ov n_i(s+1)=\ov n_i(s)+\frac{1-\alpha}{P(s)}\left[ (i-1) \ov
n_{i-1}(s)-i \ov n_i(s) \right],
\end{equation}
for $i>1$. To the families with only one individual, on the  other
hand, the   positive contribution comes  from the creation  of new
surnames with probability $\alpha$. Therefore,
\begin{equation}
\label{Simon3} \ov n_1(s+1)=\ov
n_1(s)+\alpha-\frac{1-\alpha}{P(s)}\ov n_1 (s).
\end{equation}
Note  that,  since  for  $\mu=0$  the  stochastic  process  $w(s)$
becomes trivial, overlines indicate here average over realizations
of  the  stochastic  process  by  which  each newborn's surname is
chosen.

The  system  of  equations  (\ref{Simon2})  and  (\ref{Simon3})
can be completely  solved  for  arbitrary  initial  conditions.
In fact, (\ref{Simon3})  is  an  autonomous  equation  for
$\ov n_1(s)$, whose solution reads
\begin{equation} \label{n1}
\ov n_1(s) = \frac{\alpha}{2-\alpha} (P_0+s)+ \left( n_1(0)
-\frac{\alpha}{2-\alpha} P_0 \right)
\frac{\Gamma(P_0)\Gamma(P_0+s-1+
\alpha)}{\Gamma(P_0+s)\Gamma(P_0-1+\alpha)} \, ,
\end{equation}
where  $\Gamma(z)$  is  the  gamma-function (Abramowitz \& Stegun,
1970). Then, eqn (\ref{Simon2}) can be used to recursively  obtain
$\ov  n_i(s)$  for  $i>1$.   For  long  times, $s\to \infty$, $\ov
n_1(s)$ is essentially given by  the first term in the  right-hand
side of  eqn (\ref{n1}),  which grows  linearly with  $s$, as  the
total  population.   The  second  term,  which  contains  all  the
information about the initial  condition, can be seen  to decrease
as  $s^{-(1-\alpha)}$.   In  this  limit,  the  recursion from eqn
(\ref{Simon2}) can be immediately solved. Thus, for $s\to \infty$,
we find
\begin{equation} \label{ni}
\ov n_i(s) = \frac{\alpha}{2-\alpha} \frac{\Gamma(i)\Gamma \left(
\frac{1}{1-\alpha}+2\right)}{\Gamma \left(
\frac{1}{1-\alpha}+1+i\right)} (P_0+s).
\end{equation}
The asymptotic number of families of a given size $i$ turns out to
be  proportional  to  the  total  population.  For  fixed  $s$ and
sufficiently  large  values  of  $i$,  $\ov n_i(s)$ decreases as a
power law, $\ov n_i \propto i^{-z}$, with
\begin{equation} \label{z}
z=1+\frac{1}{1-\alpha}.
\end{equation}
In  the  limit of small mutation rate, $\alpha \approx 0$, we have
$z\approx  2$.  As advanced in the Introduction, this is the value
observed  in  actual  surname  distributions. This same result was
obtained   by  Simon  (1955)  for  a  special  initial  condition.
Elsewhere  (Zanette  \&  Manrubia, 2001) we have already discussed
the  fact  that,  though  the  effects of the initial condition in
Simon's model  fade out  for long  times, transients  can strongly
depend  on  the  initial  distribution  $n_i(0)$ (see also section
\ref{num}).  In  our  case,  this  aspect  could  be  relevant  in
populations where modern surnames have appeared recently, like the
Japanese (Miyazima {\it  et al.}, 2000).  We shall return  to this
specific point later.

Equations  (\ref{Simon2})  and   (\ref{Simon3})  imply  that,   in
average,  the  total  number  of  surnames in the population grows
linearly:
\begin{equation} \label{Ns}
\ov N(s)= \sum_{i=1}^\infty \ov n_i(s) =N_0+\alpha s,
\end{equation}
with $N_0$ the  initial number of  different surnames. As  a
function of time, the number of surnames increases exponentially:
\begin{equation} \label{N0}
\ov N(t) =N_0  + \alpha P_0 [\exp( \nu  t)-1].
\end{equation}
We  have  already  pointed  out  that,  in contrast, the number of
surnames in some real  populations at present times  is decreasing
(Cavalli-Sforza \&  Feldman, 1981;  Legay \&  Vernay, 1999). Under
suitable  conditions,  adding  mortality  allows to reproduce this
particular behaviour.

\subsection{Effect of death events}
\label{analytic}

Under the action of mortality, the growth of the total  population
$P(s)$ fluctuates stochastically,  according to eqn  (\ref{Pest}),
depending  on  the  occurrence  of  death events at each evolution
step. The  evolution of  $\ov n_i(s)$  can be  implemented in  two
substeps,   as   follows.    First,   eqns   (\ref{Simon2})    and
(\ref{Simon3}) are used to calculate the intermediate values
\begin{equation}   \label{Simon2i}
\ov n_i'(s)=\ov n_i(s)+\frac{1-\alpha}{P(s)}\left[ (i-1)
\ov n_{i-1}(s)-i\ov n_i(s) \right],
\end{equation}
and
\begin{equation}    \label{Simon3i}
\ov n_1'(s)=\ov n_1(s)+\alpha-\frac{1-\alpha}{P(s)}\ov n_1 (s),
\end{equation}
and the population is updated to $P'(s)=P(s)+1$. Second, the
evolution due to mortality is applied. In  terms  of  the random
process $w (s)$ of eqn (\ref{xi}), we have
\begin{equation}   \label{Simonm}
\ov n_i(s+1)=\ov n_i'(s)+\ov {\left[ \frac{w(s)}{P'(s)} \right]}
\left[ (i+1) \ov n_{i+1}'(s)-i\ov n_i'(s) \right],
\end{equation}
and $P(s+1)=P'(s)-w(s)$ [cf. eqn (\ref{Pest})].

Naive replacement of the stochastic process $w(s)$ by its  average
value,  $w  (s)  \to  \mu$,  in  order  to  obtain a deterministic
approximation to the problem, would lead to an equation for  which
positive solutions cannot be insured.  In fact, it is possible  to
give initial conditions for such deterministic equation which lead
to negative values of $\ov  n_i(s)$ at sufficiently large $s$  and
$i$.   A satisfactory  deterministic approximation  can however be
proposed by assuming that the solution $\ov n_i(s)$ varies  slowly
both in $s$ and $i$.  This assumption is generically verified  for
large  populations  with  smooth  initial  conditions.   In   this
situation,  eqns  (\ref{Simon2i})   and  (\ref{Simonm})  admit   a
continuous approximation in terms  of real variables $z$  and $y$,
which replace  the integer  variables $s$  and $i$,  respectively.
Meanwhile,  $\ov  n_i(s)$  is  replaced  by  a continuous function
$n(y,z)$.

The  continuous  approximation  can   be  analyzed  at   different
truncation  orders,  as  discussed  in  Appendix \ref{ap2}. At the
first order, the approximate solution to eqns (\ref{Simon2i})  and
(\ref{Simonm}) reads
\begin{equation} \label{sol1}
n(y,z)=\alpha \frac{ P_0+(1-\mu)z}{1-\alpha-\mu}\
y^{-1-(1-\mu)/(1-\alpha-\mu)}
\end{equation}
for $y<y_T(z)$, and
\begin{equation}    \label{sol2}
n(y,z)= y_T^{-1} n(y/y_T(z),0)
\end{equation}
for $y>y_T(z)$, with
\begin{equation}    \label{yc}
y_T(z) =  \left(1+\frac{ 1-\mu }{P_0}z
\right)^{(1-\alpha-\mu)/(1-\mu)}.
\end{equation}

In this  first-order continuous  approximation, thus,  the average
number of families $n(y,z)$  exhibits two separated regimes.   For
large values of the size variable, $y>y_T(z)$, the distribution is
essentially determined  by the  initial condition.  At the first
evolution  steps,  where  $z\to  0$  and  $y_T \to 1$, this regime
covers  practically  all  the  domain  of  variable  $y$.  As time
elapses  and  $y_T$  grows,  however,  this  regime recedes and is
replaced  for  $y<y_T$  by  a  power-law  distribution, $n \propto
y^{-\zeta}$, with
\begin{equation} \label{zeta}
\zeta= 1+\frac{1-\mu}{1-\alpha-\mu}.
\end{equation}
Note that, as the exponent $z$ in eqn (\ref{z}), this new exponent
coincides  with  the  observed  value,  $\zeta  \approx 2$, in the
relevant limit $\alpha \approx 0$.  This result is independent  of
the  death  probability  $\mu$.  For  sufficiently long times, the
power-law  regime  will  be  observed  for all the relevant family
sizes.  All the contribution of the initial condition will  become
restricted to the domain of the largest families.

The  boundary  $y_T(z)$  between  the two regimes, eqn (\ref{yc}),
grows  as  a  power  of  the  ratio  between  the  average current
population  $P_0+(1-\mu)  z$ and the initial population $P_0$. For
$\alpha\approx  0$,  the  exponent  is practically equal to unity.
Consequently,  for  the  boundary to reach a given value $y_0$, in
such a way that all the families with sizes below $y_0$ are in the
power-law  regime,  the  total  population  must  grow by a factor
practically  equal  to  $y_0$. Taking into account eqn (\ref{Pt}),
the  position  of  the  boundary  as a function of real time reads
\begin{equation}    \label{yct}
y_T(t) = \exp[\nu (1-\alpha-\mu) t].
\end{equation}
The transient $t_0$ needed for the power-law regime to develop up
to a given size $y_0$ is therefore logarithmic, $t_0 \propto \ln
y_0$.

The  average   total  number   of  surnames   in  the   continuous
approximation is calculated by analogy with eqn (\ref{Ns}):
\begin{equation} \label{avNc}
\ov N(z)= \int_1^\infty n(y,z) dy .
\end{equation}
Using the above first-order solution  for $n(y,z)$, we find   $\ov
N(z)=N_0 +\alpha z$ or, as a function of time,
\begin{equation} \label{Nmu}
\ov N(t) =N_0 + \frac{\alpha P_0}{1-\mu} \{\exp[ \nu (1-\mu)
t]-1\} ,
\end{equation}
as  shown  in  Appendix   \ref{ap2}  [cf.  eqn  (\ref{N0})].    In
consequence,  within  this  approximation,  the  surname diversity
always  grows  exponentially.   We  point  out, however, that this
conclusion is valid for sufficiently smooth distributions and  for
$\mu<1-\alpha$,  two  necessary  conditions  for  the   continuous
approximation to  apply to  our problem.   It could  therefore  be
argued that having  a decreasing number  of surnames, as  found in
some modern developed societies (Cavalli-Sforza \& Feldman,  1981;
Legay  \&  Vernay,   1999),  requires  violation   of  the   above
conditions.   For  instance,  an  initial  condition  with   sharp
variations would violate the smoothness condition during the first
evolution stages.  A  death probability $\mu\approx 1$  would also
threaten the  validity of  the continuous  approximation. This  is
precisely  the  case  of  modern  developed  societies,  where the
population growth rate is  practically vanishing and, as  a matter
of fact,  reaches occasionally  negative values  --a situation not
described by the present model. In Section \ref{mu1} we treat  the
special  case  $\mu=1$  and  show  that,  in this limit, the total
number of surnames can in fact decrease.

Though  the  first-order  continuous  approximation gives a rather
rough description  of the  solution to  our model,  as a piecewise
function with a  moving discontinuity, it  provides a quite  clear
qualitative picture of how the solution behaves. The growth of the
power-law  regime  as  the  initial  condition  recedes is in good
qualitative  agreement  with  the  evolution observed in numerical
realizations of  the model.  A better  analytical approximation is
obtained from  the   second-order truncation.    Second-order
derivatives   would  in   fact   introduce diffusive-like
effects  in the  variable $y$,  with the consequent smoothing of
the discontinuities of the first-order approximation. The
second-order equation, however, cannot be analytically  solved
for arbitrary initial conditions. Nevertheless, it is possible  to
give an asymptotic approximation for long times, as
\begin{equation}
\label{secorder}
n(y,z) = \frac{\alpha P(z)}{1-\alpha-\mu} \left(2
\frac{1-\alpha-\mu}{1-\alpha+\mu} \right)^{\zeta-1}
y^{-1}U\left(\zeta-1,0, 2 \frac{1-\alpha-\mu}{1-\alpha+\mu}  y
\right),
\end{equation}
where $U(a,b,x)$ is the Kummer's function (see Appendix
\ref{ap2}). With respect to  the first-order approximation,  eqn
(\ref{sol1}), this solution predicts  a lower value  of $n(y,z)$
for  small $y$. For  larger  family  sizes,  however,  it behaves
as a power law, $n(y,z) \propto y^{-\zeta}$ with  exactly the
same exponent  as in eqn (\ref{zeta}).  The asymptotic behaviour
for large $y$ in  the power-law regime is therefore not modified.
It must be pointed out that, for small values of $\alpha$, the
profile of $n(y,z)$  given by eqn  (\ref{secorder}) results  to
be  quite independent  of the death probability over a
considerable range  of values of  $\mu$. Only  for  $\mu \approx
1-\alpha  \approx 1$, where the exponent $\zeta$ strongly depends
on  $\mu$, does $n(y,z)$ change  sensibly as $\mu$ is varied. We
take advantage of this feature  in Section \ref{empirical}, where
the second-order approximation is  compared with actual data of
surname distributions.

\subsection{The case of constant population, $\mu=1$}
\label{mu1}

As argued above, the limit  $\mu=1$ is relevant to the  discussion
of the evolution of surname distributions in some modern developed
societies. In fact, this limit  corresponds to the case where  the
population  growth  rate  vanishes,  a  situation which is closely
approached,  for  instance,  in  many  European countries. In this
case,  eqn  (\ref{Simonm})  is again deterministic,  with
$w(s)=1$ and $P'(s)=P_0+1$. The total population at any time is
$P(s)=P_0$.

The main difference between this case and that of $\mu<1$ is that,
now, the distribution $\ov n_i(s)$ becomes independent of time for
asymptotically large times. This feature is in agreement with  the
fact that the asymptotic distribution for $\mu<1$ is  proportional
to   the   total   population   $P(s)$.   The  asymptotic  surname
distribution  $\ov  n_i^\infty$  is  given,  for  $\mu=0$,  by the
recurrence equations
\begin{equation}
\ov  n_{i+1}^\infty=\frac{[(2-\alpha)P_0-2i(1-\alpha)]i\ov
n_i^\infty -(1-\alpha)  (P_0+1-i)(i-1)\ov
n_{i-1}^\infty}{(i+1)[P_0-(1-\alpha)(i+1)]} \end{equation} for
$i>1$, and
\begin{equation}
\ov n_{2}^\infty=\frac{[(2-\alpha)P_0-2(1-\alpha)]\ov n_1^\infty -
\alpha P_0^2}{2[P_0-2(1-\alpha)(i+1)] }.
\end{equation}
Note that, since the total population is always $P_0$, we have
$\ov n_i^\infty=0$ for $i>P_0$, as no family can be larger than
$P_0$.

The solution reveals two well-defined regimes, depending on how
the product $\alpha P_0$ compares with unity. For $\alpha P_0
>1$, the asymptotic distribution behaves as
\begin{equation}  \label{est}
\ov n_i^\infty \approx \frac{\alpha P_0}{i} (1-\alpha)^{i-1}
\end{equation}
for a vast range of family sizes. Departures from this behaviour
are found very close to $i=P_0$ only. For $i>P_0$, in fact,  the
distribution must vanish. We stress the remarkable difference
between the exponential stationary distribution (\ref{est}) and
the long-time power-law solution obtained for $\mu<1$. In this
regime, the stationary total number of surnames is
\begin{equation}\label{Nest}
\ov N^\infty \approx \frac{\alpha P_0}{1-\alpha} |\ln \alpha|.
\end{equation}

For $\alpha P_0<1$, on the other hand, the distribution behaves as
a power law,  $\ov n_i^\infty \sim  i^{-1}$, over practically  the
whole range of family sizes. Note that the exponent of this  power
law is in agreement with eqn (\ref{sol1}) in the limit $\mu\to 1$.
For $i\approx  P_0$, however,  the distribution  deviates from the
power law and exhibits a  sharp peak. In the limit  $\alpha\to 0$,
the distribution becomes an isolated peak at $i=P_0$, namely, $\ov
n_i^\infty =0$ for  $i<P_0$ and $\ov  n_i^\infty =1$ for  $i=P_0$.
Therefore, the  total number  of surnames  is $\ov  N^\infty = 1$.
This special solution  describes the well-known  case of a  closed
population with  no surname  mutations where,  by random  drift, a
single   surname   survives   for   asymptotically   large   times
(Cavalli-Sforza \&  Feldman, 1981).  Notice also  that this  limit
corresponds to the birth-death model introduced by Moran (1962) to
study  probabilities  of  fixation  of  alleles  when  generations
overlap.

We conclude  that, for  a given  population $P_0$,  the asymptotic
number of surnames can be  very small if the mutation  probability
$\alpha$  is,  in  turn,  small  enough. Consequently, in a steady
population with  many surnames  at the  initial stage  and with  a
sufficiently low mutation probability, the number of surnames will
decrease towards the stationary value as time elapses.   Numerical
simulations, discussed in  detail in the  next section, show  that
for  $\mu$  just  below  unity  and  small  $\alpha$,  an  initial
transient where the number  of surnames decreases temporarily  can
be observed. Since $\mu<1$, however, the population grows steadily
and, as a consequence of mutations, $\ov N$ will also increase  in
the long run. The situation in modern developed societies is  that
the  population  growth  rate  has  been constantly decreasing, to
reach values around zero at present times.  Starting from a  state
with a  wealth of  surnames --due  to the  combined effect,  a few
centuries  ago,  of  a  high  population  growth  and the frequent
appearance of new  surnames-- these societies  have now reached  a
regime  of  almost  stationary  population  where  the  number  of
surnames decreases. This situation would be reverted if the growth
rate could  be maintained  above zero  during substantial periods.
Presently, the only mechanism acting in this direction seems to be
immigration,  which  is  in  addition  an  effective source of new
surnames.

\section{Numerical results}
\label{num}

In this section  we present results  of numerical realizations  of
our model, in order to compare with the analytical  approximations
presented  in  Section  \ref{analytic},  and  to  illustrate   the
behaviour  within  the  regimes  where  such approximations do not
hold. Emphasis is put on  the role of the initial  conditions, the
duration  of  transients,  and  the  evolution  of  the  number of
different surnames. Also, we discuss actual surname  distributions
of modern populations in the light of our results.

\subsection{Role of initial conditions}
\label{initcond}

First,  it  is  worthwhile  to  illustrate  how  the  shape of the
distribution  $\ov  n_i(s)$  depends  on  the  initial   condition
$n_i(0)$  [see  also  Zanette  \&  Manrubia  (2001)]. We focus the
attention on  initial distributions  of the  form $n_i(0)=N_0$ for
$i=i_0$,  and  $n_{i}(0)=0$  for  $i\neq  i_0$, in which there are
$N_0$ families  of equal  size formed  by $i_0$  individuals each.
Consequently,  the  initial  population  is  $P_0=i_0N_0$.  In the
following,  we  denote  such  an  initial  condition  by  the pair
$(i_0,N_0)$.     Figure  3   shows   four   normalized
distributions of family sizes,
\begin{equation} \label{p_i}
p_i(s)= \frac{\ov n_i(s)}{\sum_i\ov n_i(s)},
\end{equation}
obtained both  from the  numerical realization  of the  stochastic
birth-death  model  and  from   the  iterative  solution  of   the
deterministic eqns (\ref{Simon2}) and (\ref{Simon3}), for $\mu=0$.
We find  a very  good agreement  between both  methods and, at the
same  time,  clearly  realize  the  relevant  role  of the initial
condition in determining the profile of the distribution for large
family sizes. These  same features are  found for other  values of
the death probability.

The solution of the  first-order approximation to our  model, eqns
(\ref{sol1}) to  (\ref{yc}), predicts  that in  the zone  of small
family sizes --i.e. for $y<y_T$ in the continuous variables--  the
only  dependence  on  the  initial  condition  appears through the
quantity $P_0$. This means  that the distribution of  family sizes
is not sensitive to a variation of $i_0$ and $N_0$ as far as their
product  is  kept  constant.  Note  that  the three cases $(4,5)$,
$(1,20)$, and $(20,1)$ of  Fig. 3 share the  same value of
$P_0=20$.  For  the  parameters  of  the figure, the crossover
between the regions of small and large family sizes, given for the
continuous  variables   by  eqn   (\ref{yc}),  should   occur   at
$i_T\approx 470$. Though two of the distributions indeed have  the
same  profile  --with  the  expected  power-law decay-- up to that
value,  the  distribution  corresponding  to the initial condition
$(1,20)$ deviates considerably below $i_T$. This deviation can  be
attributed  to  the  fact  that  the  initial  condition  $(1,20)$
corresponds to a quite singular distribution, with a high peak  at
$i=1$.  A  continuous  approximation  for  such  a distribution is
arguably  expected  to  give  a  poor  description  of  the   real
situation.

\begin{figure}
\centerline{\psfig{file=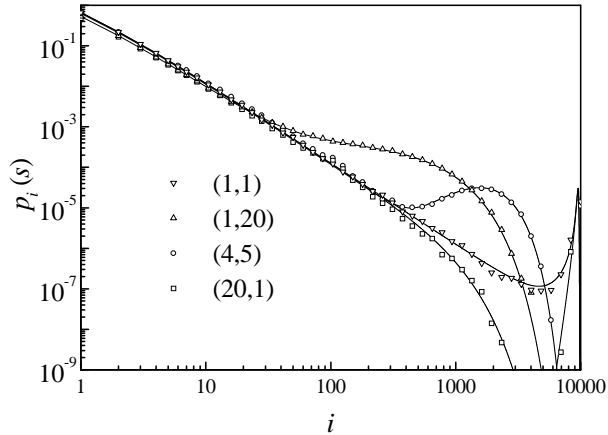,width=9.5 cm}}
\caption{Effect of the initial condition on the tail of the
surname distribution. The normalized distribution $p_i(s)$ is
shown for $s=10^4$, $\mu=0$, $\alpha=10^{-2}$, and four different
initial conditions, as specified in the legend. Symbols stand for
numerical simulations of the birth-death stochastic model,
averaged over $2000$ independent realizations. Curves correspond
to the numerical solution of the deterministic equations
(\ref{Simon2}) and (\ref{Simon3}).}
\label{fig:p0}
\end{figure}

\subsection{Computational measurement of transients}

Equation (\ref{yc}) defines, within the first-order approximation,
the boundary  that separates  the asymptotic  regime and  the zone
dominated by  the initial  condition. Alternatively,  for a  fixed
family size, it can be used to determine the transient before  the
asymptotic distribution builds up at  that size. In order to  test
this aspect of the  first-order approximation, we have  devised an
independent computational  method to  evaluate the  point at which
the  cut-off  between  the  two  regimes  actually  takes   place.
According  to  eqns  (\ref{sol1})  and (\ref{p_i}), the asymptotic
normalized  distribution  of  family  sizes  is $p_i^\infty \equiv
p_i(s\to  \infty)  \approx  (\zeta-1)  i^{-\zeta}$.  In  numerical
realizations, the  distribution is  expected to  adopt values very
close to $p_i ^\infty$ for large $i$ and large enough $s$. In  the
simulations,  we  fix  a  certain  family  size $i_T$ and stop the
calculation at the step $s_T$ when the measured value of $p_{i_T}$
satisfies,  for  the  first  time,  $|p_{i_T}  - p_{i_T}^\infty| <
\Delta$. The results presented in the following correspond to  the
choice $i_T=10^2$ and $\Delta=10^{-1}$. Since, as shown in Section
\ref{initcond}, the agreement between the numerical simulations of
the stochastic birth-death model and the iterative solution of the
corresponding average equations is  very good, we use  this second
description to measure the transient $s_T$.

We  focus  the  analysis  on  the  dependence  with  the   initial
conditions, and  keep the  values of  $\alpha=10^{-2}$ and $\mu=0$
fixed. As far as $\alpha\approx 0$, the results are  qualitatively
the same for other values of $\mu$. We consider initial conditions
of the  form $(i_0,N_0)$,  as defined  in Section  \ref{initcond}.
Figure 4  shows the  transient $s_T$  measured from the numerical
solution  to  eqns  (\ref{Simon2})  and  (\ref{Simon3})
according  to  the  criterion  introduced  above, as a function of
$i_0$ and $N_0$. On the one hand, for small $N_0$ and, especially,
for  small  $i_0$,  $s_T$  varies  quite irregularly. On the other
hand,  we  see  that  for  larger  values  of $i_0$ and $N_0$, say
$i_0>10$  and  $N_0>5$,  $s_T$  exhibits  a  well-defined   linear
dependence with both parameters. In this regime, the transient  is
well approximated by
\begin{equation} \label{stn}
s_T= a i_0 N_0.
\end{equation}
\begin{figure}
\label{fig:i0}
\centerline{\psfig{file=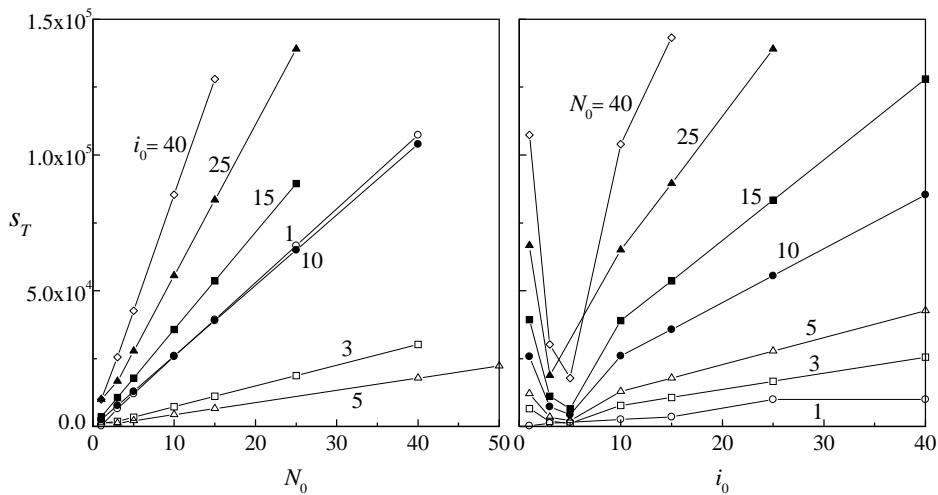,width=\textwidth}}
\caption{Numerically measured transient $s_T$ as a function of the
initial  parameters  $i_0$  and  $N_0$,  for  $\alpha=10^{-2}$ and
$\mu=0$. The parameters that  define the transient (see  text) are
$i_T=10^2$ and $\Delta=10^{-1}$.}
\end{figure}

Linear fitting of $s_T$ as a  function of $i_0$ and $N_0$ for  the
data shown in Fig. 4 yields $a=237\pm 4$. Note that eqn
(\ref{stn}) implies that, for large $i_0$ and $N_0$ the  transient
length is  proportional to  the initial  population $P_0=i_0 N_0$.
This  is  to  be  compared  with  the  analogous result within the
first-order approximation, eqn (\ref{yc}), which implies
\begin{equation} \label{sta}
s_T = \frac{P_0}{1-\mu} \left[ i_T^{(1-\mu) / (1-\alpha-\mu)}-1
\right].
\end{equation}
The  first-order  approximation  predicts  then  a proportionality
between $s_T$ and $P_0$ --independently of the values of  $\alpha$
and $\mu$-- in full agreement with the numerical results. We point
out, however, that the corresponding proportionality  coefficients
cannot be directly compared. In  fact, the coefficient $a$ in  eqn
(\ref{stn}) is evaluated following the computational definition of
the transient, and thus  depends on the parameter  $\Delta$, which
has no correlate in  the analytical approach. Note,  nevertheless,
that for $\alpha\approx 0$ and $\mu=0$ eqn (\ref{sta}) predicts  a
coefficient approximately  equal to  $i_T =100$,  which is  of the
same order as the numerical result.

\subsection{Evolution of surname diversity}

As discussed in Section \ref{model}, the evolution of the  average
number $\ov N(s)$ of different surnames is driven in our model  by
two  competing  mechanisms.  The  diversity  increases  due to the
appearance of new surnames at rate $\alpha$, and surnames borne by
single individuals disappear when the individual in question dies.
Within the first-order continuous approximation to our model,  for
$\mu<1$,  the  average  number  of  different  surnames  increases
linearly  with  $s$  as  $\ov  N(s)=N_0+\alpha  s$  (see   Section
\ref{analytic}).  In  this  approximation,  the  death probability
$\mu$ plays a role in the variation of diversity as a function  of
time, eqn (\ref{Nmu}).  On the one  hand, increase of  the surname
diversity in a steadily  growing population is generally  expected
when new surnames are created at a constant rate. On the other, it
is possible  to conceive  special situations  where the  number of
different  surnames  should  temporarily  decrease,  violating the
first-order approximation. Imagine, for instance, that the initial
population  consists  of  an  ensemble  of  families with only one
individual each. For moderate values of $\mu$ and small  $\alpha$,
the initial  stage would  be characterized  by the  death of  some
individuals, with the consequent disappearance of their  surnames,
and no significative appearance of new surnames.  Since the  total
population grows,  however, $\ov  N(s)$ will  eventually attain  a
minimum and, from then on, will increase.

To  illustrate   this  situation,   we  numerically   solve   eqns
(\ref{Simon2i})  to  (\ref{Simonm})  for  the  initial   condition
$(1,20)$, with  $20$ families  of one  individual each.  Note that
this is the  initial condition for  which we detected  the largest
deviations  from   the  first-order   approximation  in    Section
\ref{initcond}.  Figure 5  shows  the evolution of $\ov N(s)$ for
different  values  of  $\mu$.  As expected, an initial transient
where the surname diversity drops is found for  $\mu>0$. The
transient  is longer  and the  minimum in  $N(s)$ is deeper as
$\mu$ grows. In all cases, however, the subsequent growth of  $\ov
N(s)$ is clearly seen. Note that the slope of this growth  depends
slightly  on  $\mu$,  a  feature  not predicted by the first-order
continuous approximation.

These long transients where (for large $\mu$ and suitable  initial
conditions)  the  number  of  different  surnames  is  expected to
decrease  could  be   relevant  to  the   description  of   modern
populations with declining  diversity (Cavalli-Sforza \&  Feldman,
1981;  Legay  \&  Vernay,  1999).  A realistic description of this
situation should however take into account that, in recent  times,
the relative values of birth and death rates in actual populations
have considerably changed. The general trend to population  growth
observed  worldwide  in  the  nineteenth  century  has by now been
reversed in  many developed  areas, such  as in  Europe, where the
total population  is practically  stationary. In  our model,  this
corresponds to an increment in the value of $\mu$ to values  close
to unity. In particular, the death probability must be allowed  to
vary with time.  Additionally, if the  description is expected  to
encompass the  periods where  the appearance  of new  surnames was
frequent --specifically, the Middle  Ages in the case  of Europe--
$\alpha$ should also change during the evolution.

\begin{figure}
\centerline{\psfig{file=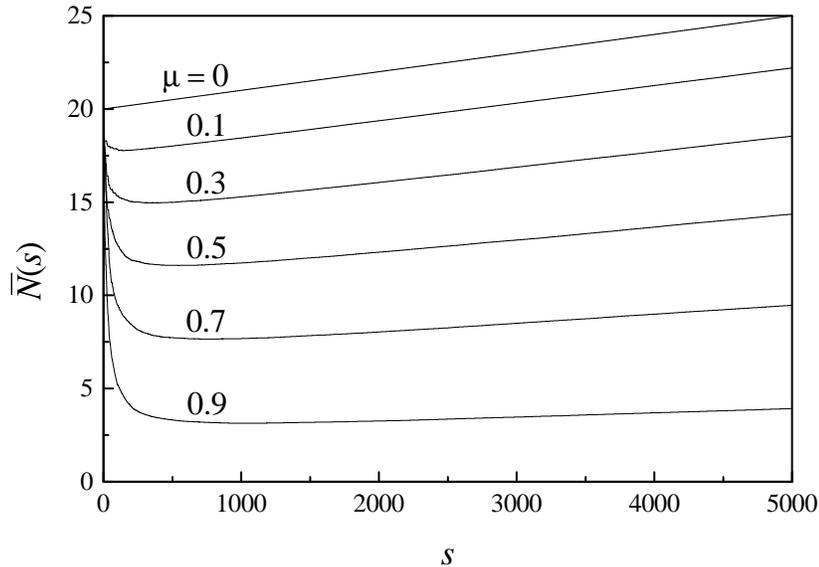,width=12 cm}}
\caption{Evolution of the average number of different surnames for
the  initial   condition  $(1,20)$,   with  $\alpha=10^{-3}$   and
different values of the death probability $\mu$.}
\label{fig:N}
\end{figure}

As a qualitative demonstration of the effect of varying the  model
parameters with time, we focus the attention on a change of  $\mu$
from  a  low  value  to  a  high  value. For the initial condition
$(1,20)$ we fix  $\mu=0$ during the  first $s_0$ evolution  steps.
Then, $\mu$ is  left to increase  linearly with $s$,  such that it
reaches  unity  after  $s_1$  additional  steps.  In the numerical
calculations shown in Fig. 6 we fix $s_1= 10^3$  and consider
several  values  of  $s_0$.  For  $s<s_0$, the number of
different surnames increases (cf. Fig. 5). Then, as  the death
probability grows, $\ov  N(s)$ attains a maximum  and begins to
decrease.  For asymptotically  large times,  it is  expected to
approach a stationary value, as predicted for the case $\mu=1$  in
Section \ref{mu1}.  Interestingly, $\ov  N(s)$ responds  faster to
the effects of  growing $\mu$ for  smaller values of  $s_0$, where
the  contribution  of  the  initial  condition  is still important
during  the  variation  of  the  death probability. The asymptotic
surname distribution seems to be,  in this sense, quite robust  to
the action of varying $\mu$. This feature should be related to the
fact that the asymptotic distribution is rather insensitive to the
value of $\mu$, as discussed at the end of Section \ref{analytic}.

\begin{figure}
\centerline{\psfig{file=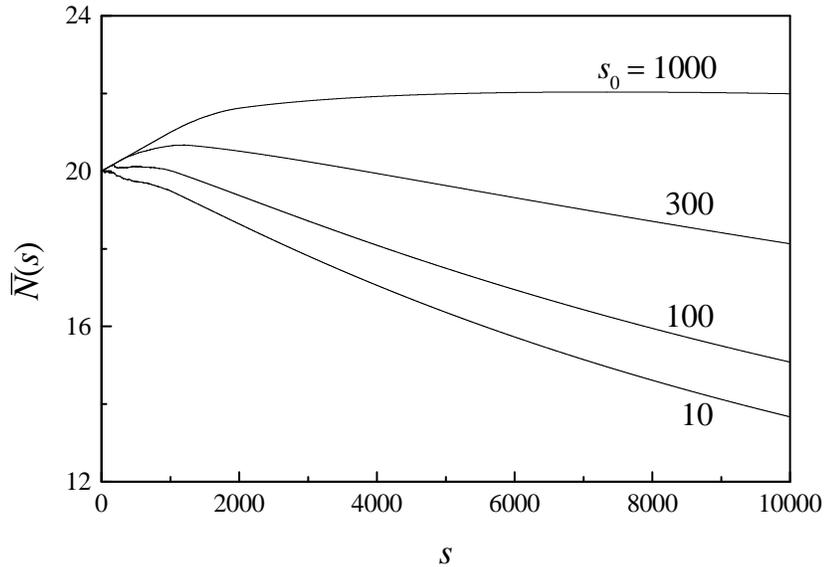,width=12 cm}}
\caption{Evolution of the diversity of surnames under variation
of the death probability $\mu$, for $\alpha=10^{-3}$. See main
text for further explanations.}
\label{fig:muvar}
\end{figure}

\section{Comparison of the model with field data}
\label{empirical}

Finally, we compare some of the analytical results for our  model,
derived under  several approximations  in Section  \ref{analytic},
with actual data from  three modern populations. Specifically,  we
focus  the  attention  on  the  second-order  approximation,   eqn
(\ref{secorder}), for the  asymptotic distribution of  surnames in
the range of small family  sizes, where the effect of  the initial
condition is negligible.  In fact, it  is virtually impossible  to
extract information on the distribution of surnames in  historical
times  from  the  data  presently  available.  We  recall that eqn
(\ref{secorder})  correctly  describes  the  asymptotic  power law
$n(y,z) \propto y^{-\zeta}$ with $\zeta \approx 2$, as observed in
real  data.  The  validity  of  our  second-order approximation is
therefore to be especially evaluated  in the region of very  small
family sizes, where the distribution differs from the power law.

Our three data sets were obtained from surname counts in telephone
books.  They  correspond  to  (i)  the  almost  350,000  different
surnames  of  the  whole  1996  Argentine telephone book, (ii) the
6,400 surnames beginning by A  in the 1996 Berlin telephone  book,
and (iii) the surnames  in five Japanese cities,  with populations
ranging between  $2\times 10^3$  and $2\times  10^5$, reported  by
Miyazima {\it  et al.}  (2000). Let  us point  out that  the three
populations involved here have considerably different  demographic
history.  The  modern   Argentine  population  has   predominantly
European ancestors, who immigrated mainly in the period  1880-1915
and just after  the World War  II. Their surname  distribution has
therefore to be considered as a combined sample from the countries
that contributed  the immigrants.  From the  times of  the largest
immigration waves to the present,  both the birth and death  rates
have  substantially  changed.  As   for  Berlin,  this  city   was
practically  abandoned  in  the  late  stages  of World War II and
subsequently repopulated with a mixture of the ancient inhabitants
and newcomers  from other  cities of  Germany. In  this case,  the
surname distribution has consequently resulted from a  combination
of several  German regions.  The modern  population in  Berlin, in
addition, presents the  particularity of having  been artificially
separated into  two practically  immiscible groups  during several
decades, ending in 1989.  Since European surnames appeared  mostly
during the Middle  Ages (Legay \&  Vernay, 1999), the  populations
that  contributed  surnames  to  both  Argentina  and  Berlin  are
expected to have developed the asymptotic surname distribution  on
a substantial range of family sizes. This situation contrasts with
the case of  modern Japanese surnames,  that were originated  some
$120$ years ago (Miyazima {\it et al.}, 2000). In Japan, moreover,
the contribution of immigration in the relevant period should have
been considerably less important than in Berlin and Argentina.

\begin{figure}
\centerline{\psfig{file=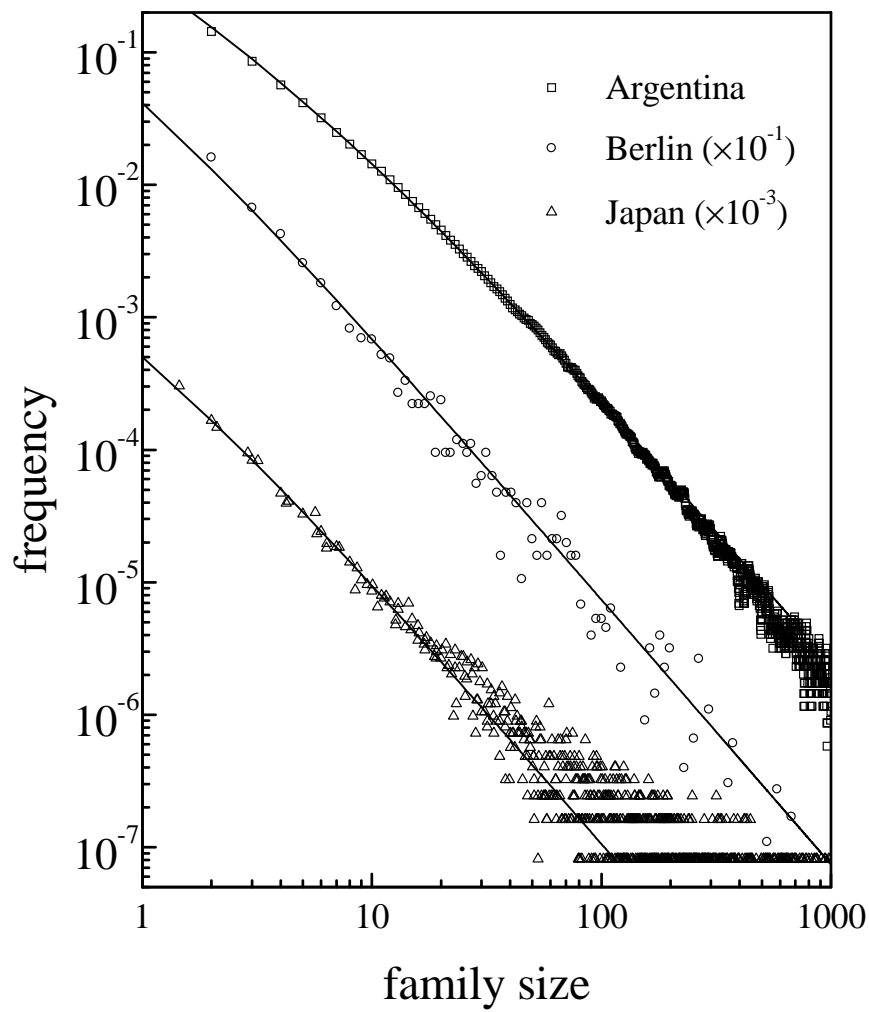,width=12 cm}}
\caption{Frequency of  family names  as a  function of  the family
size for three  modern populations. Dots  correspond to real  data
from the sources quoted in the text.  Curves stand for the fitting
of each data set  with eqn (\ref{secorder}), taking  $\mu=0.7$ for
Argentina and  $\mu=0.2$ for  Berlin and  Japan. For  clarity, the
data for Berlin and Japan have been shifted by a constant factor.}
\label{fig:XX}
\end{figure}

Explicit evaluation  of eqn  (\ref{secorder}) requires  fixing the
values of  the mutation  rate $\alpha$  and the  death probability
$\mu$. We know that the mutation rate, i.e. the probability that a
new individual acquires  a surname not  previously present in  the
population, is very low for  any modern society. Since, as  far as
$\alpha\neq 0$, the limit of eqn (\ref{secorder}) for $\alpha  \to
0$  is  well  defined,  fixing  any  sufficiently  small value for
$\alpha$  gives  a  correct   description  of  $n(y,z)$.  In   our
comparison with real data, we have taken $\alpha=10^{-3}$. As  for
the  death  probability  $\mu$,  unfortunately,  we  have found it
impossible  to  fix  a  reliable  value. For all real populations,
mortality has considerably changed during the periods relevant  to
the  evolution  of  surname  frequencies.  Moreover,  since  $\mu$
measures the relative  frequency of death  and birth events  --the
latter  also  including,  in  real  populations,  arrival  of  new
individuals  by   immigration--  an   evaluation  of   the   death
probability  should  also  involve   a  detailed  description   of
immigration effects. On the other hand, as discussed at the end of
Section \ref{analytic}, the asymptotic profile of $n(y,z)$,  given
by eqn (\ref{secorder}), is  practically independent of the  death
probability as far  as $\mu$ is  not close to  unity. We therefore
decided to fit the  field data with eqn  (\ref{secorder}) allowing
$\mu$  to  vary  in  order  to  get  the best approximation in the
relevant zone of small family sizes.

Figure  7   shows  the   three  data   sets  and    the
corresponding fittings  with eqn  (\ref{secorder}). Fittings  have
been optimized  in the  domain of  small family  sizes, where  our
analytical  approximation  is  expected  to  hold.  In the case of
Argentina, the agreement with real data is excellent up to  family
sizes above $i=100$.  Only for $i>200$  a systematic deviation  is
observed, where the analytical result overestimates the frequency.
According  to  our  discussion  in  Section  \ref{analytic},  this
deviation would be the remnant contribution of initial conditions.
For the Berlin  data, the statistics  are poorer.   However, it is
clear that the  analytical approximation fits  the data with  good
precision  in  the  whole  range  shown  here.  In the case of the
Japanese data the fitting is very good for relatively small family
sizes,  $i<10$,  but,  on  the  other  hand, noticeable systematic
deviations appear for $i>20$. This agrees with our expectation  on
the effect of initial conditions. In the $120$ years elapsed  from
the appearance of modern Japanese family names, the population  in
the cities from which the present data were obtained has increased
by a factor of, at most, order $10$.  Therefore, according to  eqn
(\ref{yc}),   initial   conditions   should   contribute   to  the
distribution for  relatively small  family sizes,  $i>10$, just as
observed.

\section{Discussion}
\label{disc}

We have analysed a birth-death model with overlapping generations,
in order to study several statistical properties of monoparental
inheritance in large populations. We have focused the analysis on
a cultural trait, namely, the distribution of surnames, taking
advantage of the availability of big corpora of real data.
However, the model applies generally to biological traits
associated with non recombining  alleles. Our model has two
parameters, namely, the probability that a new individual carries
a surname not present in the population, $\alpha$, and the
average number of death events per birth $\mu$. For any value of
$\alpha>0$ the system attains a broad, stationary distribution of
surname diversity. If $\mu<1$ the total population grows
exponentially in time and so does the total number of different
surnames. The marginal case $\mu=1$ corresponds to constant total
population and, on average, to constant surname diversity for
asymptotically large times.

Our analytical results for the stationary distribution of
surnames frequency are in good agreement with field data for
modern human populations in different countries. Through an
analysis of the transient time required for this distribution to
reach its asymptotic shape, we have shown that some deviations
observed in real data might actually reflect the composition of
the founder population. This result has implications in the study
of polyphyletism. Indeed, if the same surname can have multiple
origins and thus the individuals carrying it are not always
phylogenetically related, this will affect the shape of the
surname distribution. In particular, it is not difficult to
estimate the time when surnames originated in a population (using
historical records) or, in the biological counterpart, when a
mutant allele first appeared (through molecular clock analysis).
Then, the approximate cut-off until which the stationary
distribution follows the asymptotic shape can be calculated. If
the distribution underestimates the frequency of values larger
than the cut-off, the system is mainly polyphyletic. If it
overestimates that frequency, then the simultaneous appearance of
many individual carrying different surnames took place in the
past.

The strong resemblance between the cultural inheritance of the
surname and the biological process in which non recombining
neutral alleles are passed to offspring has justified to apply
results from field data in the former case to the latter (Barrai
{\it et al.}, 1996). In the few cases where data on genetic
diversity was available, it was possible to retrieve information
on past populations by comparing both sets of data (Sykes \&
Irven, 2000). A specific example comes from the small island of
Tristan da Cunha, where the fact that its $300$ inhabitants
represent only seven surnames and five mitochondrial lineages
reflects without doubt the small size of the founder population
(Soodyall {\it et al.}, 1997).

Our results could be applied to population genetics under some
hypotheses. Indeed, we are assuming that the number of different
alleles at a given locus is practically infinite (this is
analogous to the assumption made by Kimura \& Crow (1964), in
their model of infinitely many alleles), since the possibility of
backward mutations is discarded. Nonetheless, this factor could
be accounted for just by lowering the value of $\alpha$, because
this process implies that the ``new'' mutant is in fact identical
to one of the forms present in the population. Since, as long as
$\alpha$ is small, our results are not sensibly modified, we
could also work with a finite but large number $R$ of different
surnames (equivalently, large genetic polymorphism), and use the
same model as long as the current diversity is lower than $R$. We
have also assumed that the death rate is constant during the life
time of individuals, while it is known that the life expectancy
depends not only on the age of individuals, but also varies as a
function of time (Vaupel {\it et al.}, 1998). A more realistic
model of human inheritance could be constructed by taking into
account the variation of $\mu$ along the lifetime of each
individual.

The evolution of surname diversity follows truly neutral
evolution: family names do not fulfill any practical purpose but
identifying the lineage of each individual. They cannot be
selected for through natural selection. Hence, as we have shown,
their statistical distribution closely follows the predictions of
a neutral model of monoparental inheritance. It would be
interesting to test our theoretical results against the genetic
diversity of a large population sample. Unfortunately, however,
data on frequencies of allelic polymorphisms are still scarce to
carry out a massive study like the one presented here for actual
surname distributions. Moreover, one should make sure that the
different haplotypes in the sample do not confer any advantage to
the individuals, in which case positive feedbacks and deviations
from neutral statistics would be expected. Different tests have
been proposed in the literature to detect deviations from
neutrality (Nielsen, 2001; Fu \& Li, 1993), and a number of
neutral haplotypes have been positively identified (S\'anchez
Mazas {\it et al.}, 1994; Stenoien, 1999). Hopefully, enough data
will be available in a near future to calculate reliable
diversity distributions in human populations.

\section*{Acknowledgements}

We thank M. Montemurro for kindly suplying the whole surname
count from the 1996 Argentine telephone book.

\section*{References}

\noindent
{\sc Abramowitz, M. \& Stegun, I. A.} (1970). {\it Handbook of
Mathematical Functions}. New York: Dover. \vspace{5 pt}

\noindent
{\sc Barrai, I., Scapoli, C., Beretta, M., Nesti, C., Mamolini, E.
\& Rodriguez - Larralde, A.} (1996). Isonymy and the genetic
structure of Switzerland. I. The distribution of surnames. {\it
Ann. Hum. Biol. \bf 23}, 431-455. \vspace{5 pt}

\noindent
{\sc Burlando, B.} (1990). The fractal dimension of taxonomic
systems.  {\it J. theor. Biol. \bf 146}, 99; (1993). The fractal
geometry of evolution. {\it J. theor. Biol. \bf 163},
161.\vspace{5 pt}

\noindent
{\sc Cavalli-Sforza, L. L. \& Feldman, M. W.} (1981). {\it
Cultural Transmission and Evolution: A Quantitative Approach}.
Princeton:  Princeton University Press.\vspace{5 pt}

\noindent
{\sc Cavalli-Sforza, L. L., Feldman, M. W., Chen, K. H. \&
Dornbusch, S.M.} (1982). Theory and observation in cultural
transmission.  {\it Science \bf 218}, 19-27.\vspace{5 pt}

\noindent
{\sc Consul, P. C.} (1991). Evolution of surnames. {\it Int.
Stat. Rev. \bf 59}, 271-278.\vspace{5 pt}

\noindent
{\sc Darwin, C.} (1871). {\it The Descent of Man and Selection in
Relation to Sex.}   London:  Murray.\vspace{5 pt}

\noindent
{\sc Dewdney, A. K.} (1986). Computer recreations: Branching
phylogenies of the Paleozoic and the fortunes of English family
names. {\it Sci. Am. \bf 254}, 12-16.\vspace{5 pt}

\noindent
{\sc Fisher, R.A.} (1922). On the dominance ratio. {\it Proc.
Roy. Soc. Edin. \bf 42}, 321-341.\vspace{5 pt}

\noindent
{\sc Fox, W.R. \& Lasker, G.W.} (1983). The distribution of
surname frequencies. {\it Int. Stat. Rev. \bf 51},
81-87.\vspace{5 pt}

\noindent
{\sc Fu, Y.X.  \& Li, W.H. } (1993). Statistical tests of
neutrality of mutations. {\it Genetics \bf 133},
693-709.\vspace{5 pt}

\noindent
{\sc Gale, J.S.} (1990). {\it Theoretical population genetics.}
Unwin Hyman, London.\vspace{5 pt}

\noindent
{\sc Galton, F. \& Watson, H.W.} (1874). On the probability of
the extinction of families. {\it J. Roy. Anthropol. Inst. \bf 4},
138-144.\vspace{5 pt}

\noindent
{\sc Gomes, M. A. F., Vasconcelos, G. L., Tsang, I. J. \& Tsang,
I. R.} (1999). Scaling relations for diversity of languages. {\it
Physica A \bf 271}, 489-495.\vspace{5 pt}

\noindent
{\sc Greenberg, J. H.} (1992). Preliminaries to a systematic
comparison between biological and linguistic evolution. In: {\it
The evolution of human languages} (Hawkins J. A. and  Gell-Mann
M., eds) Reading: Addison-Wesley.\vspace{5 pt}

\noindent
{\sc Haldane, J. B. S.} (1927). A mathematical theory of natural
and artificial selection. Part V: Selection and mutation. {\it
Proc. Camb. Phil. Soc. \bf 26}, 838-844.\vspace{5 pt}

\noindent
{\sc Harris, Th. E.} (1963). {\it The theory of branching
processes.} Berlin: Springer.\vspace{5 pt}

\noindent
{\sc Hull, D. M.} (1998). A reconsideration of Galton's problem
(using a two-sex population). {\it Theor. Pop. Biol. \bf 54},
105-116.\vspace{5 pt}

\noindent
{\sc Islam, M. N.} (1995). A stochastic model for surname
evolution. {\it Biom. J. \bf 37}, 119-126.\vspace{5 pt}

\noindent
{\sc Karlin, S. \& McGregor, J.} (1967). The number of mutant
forms maintained in a population. {\it Proc. fifth Berkeley Symp.
Math. Stat. Prob. \bf 4}, 415-438.\vspace{5 pt}

\noindent
{\sc Kimura, M. \& Crow, J.F.} (1964). The number of alleles that
can be maintained in a finite population. {\it Genetics \bf 49},
725-738.\vspace{5 pt}

\noindent
{\sc Lange, K.} (1981). Minimum extinction probability for
surnames and favorable mutations. {\it Math. Biosci. \bf 54},
71-78.\vspace{5 pt}

\noindent
{\sc Legay, J. M. \&  Vernay, M.} (1999). L' instructive histoire
des noms de familie.  {\it Pour la Science} n. 255,
58-65.\vspace{5 pt}

\noindent
{\sc Lotka, A. J.} (1931). Population analysis--the extinction of
families--I.  {\it Washington Acad. Sci. \bf 21}, 377-380; {\it
ibid.} 453-459.\vspace{5 pt}

\noindent
{\sc Miyazima, S., Lee, Y., Nagamine, T. \& Miyajima, H.} (2000).
Power-law distribution of family names in Japanese societies.
{\it Physica A \bf 278}, 282-288.\vspace{5 pt}

\noindent
{\sc Moran, P.A.P.} (1962). {\it The statistical processes of
evolutionary theory.} Clarendon Press, Oxford.\vspace{5 pt}

\noindent
{\sc Nielsen, R.} (2001). Statistical tests of selective
neutrality in the age of genomics. {\it Heredity \bf 86},
641-647.\vspace{5 pt}

\noindent
{\sc Panaretos, J.} (1989). On the evolution of surnames. {\it
Int. Stat. Rev. \bf 57}, 161-167.\vspace{5 pt}

\noindent
{\sc Pielou, E. C.} (1969). {\it An introduction to mathematical
ecology.} New York: Wiley.\vspace{5 pt}

\noindent
{\sc Poore, M. E. D.} (1968). Studies in Malaysian rainforest.
{\it Ecology \bf 56}, 143-196.\vspace{5 pt}

\noindent
{\sc Rannala, B.} (1997). Gene genealogy in a population of
variable size. {\it Heredity \bf 78}, 417-423.\vspace{5 pt}

\noindent
{\sc Ruhlen, M.} (1992). An overview of genetic classification.
In: {\it The evolution of human languages} (Hawkins, J.A. and
Gell-Mann M., eds) Reading: Addison-Wesley.\vspace{5 pt}

\noindent
{\sc S\'anchez Mazas, A., Butler Brunner, E., Excoffier, L.,
Ghanem, N., Salem, M.B., Breguet, G., Dard, P., Pellegrini, B.,
Tikkanen, M.J., Langaney, A., Lefranc, G., \& Butler, R.} (1994).
New data for AG haplotype frequencies in caucasoid populations
and selective neutrality of the AG polymorphism. {\it Human
Biology \bf 66}, 27-48.\vspace{5 pt}

\noindent
{\sc Simon, H. A.} (1955). On a class of skew distribution
functions. {\it Biometrika \bf 42}, 425-440.\vspace{5 pt}

\noindent
{\sc Soodyall, H., Jenkins, T., Mukherjee, A. du Toit, E.,
Roberts, D.F. \& Stoneking, M.} (1997). The founding mitochondrial
DNA lineages of Tristan da Cunha Islanders. {\it Am. J. Phys.
Anthropol. \bf 104}, 157-166.\vspace{5 pt}

\noindent
{\sc Sol\'e, R. V. and Alonso, D.} (1998). Random walks, fractals,
and the origin of rainforest diversity. {\it Adv. Complex Systems
\bf 1}, 203-220.\vspace{5 pt}

\noindent
{\sc Stenoinen, H.K.} (1999). Are enzyme loci selectively neutral
in haploid populations of nonvascular plants? {\it Evolution \bf
53}, 1050-1059.\vspace{5 pt}

\noindent
{\sc Sykes, B. \& Irven, C.} (2000). Surnames and the Y
chromosome. {\it Am. J. Hum. Genet. \bf 66}, 1417-1419.\vspace{5
pt}

\noindent
{\sc van Kampen, N. G.} (1981). {\it Stochastic Processes in
Physics and Chemistry.} Amsterdam: North-Holland.\vspace{5 pt}

\noindent
{\sc Vaupel, J. W., Carey, J. R., Christensen, K., Johnson, T. E.,
Yashin, A. I., Holm, N. V., Iachine, I. A., Kannisto, V.,
Khazaeli, A. A., Liedo, P., Longo, V. D., Zeng, Y., Manton, K. G.
\& Curtsinger, J. W.} (1998). Biodemographic trajectories of
longevity. {\it Science \bf 280}, 855-860.\vspace{5 pt}

\noindent
{\sc Yasuda, N., Cavalli-Sforza, L. L., Skolnick, M. \& Moroni,
A.} (1974). The evolution of surnames: An analysis of their
distribution and extinction. {\it Theor. Pop. Biol. \bf 5},
123-142.\vspace{5 pt}

\noindent
{\sc Zanette, D. H.} (2001). Statistical regularities in the
taxonomic classification of human languages. {\it Adv. Complex
Systems \bf 4}, 281-286.\vspace{5 pt}

\noindent
{\sc Zanette, D. H. \& Manrubia, S.C.} (2001). Vertical
transmission of culture and the distribution of family names.
{\it Physica A \bf 295}, 1-8.\vspace{5 pt}

\noindent
{\sc Zipf, G. K.} (1949) {\it Human Behavior and the Principle of
Least Effort}. Cambridge: Addison-Wesley.

\appendix

\section{Average evolution of the total population}
\label{ap1}

As explained in the main text, at each step, the total  population
$P(s)$  in  our  model  either  increases  by $1$ with probability
$1-\mu$,  due  to  the  occurrence  of  a birth event and no death
event,  or  remains  constant  with  probability $\mu$, due to the
occurrence  of  both  a  birth  and  a  death  event.    Equations
(\ref{Pest}) and (\ref{xi}) quantify the corresponding  stochastic
process.

Since the total population changes during the evolution, the  time
interval  $\delta  t$  to  be  associated  with  each step --which
corresponds to the interval between consecutive births-- must also
change. In fact, the birth frequency is proportional to the  total
population,  so  that  $\delta  t$  is  inversely  proportional to
$P(t)$. We write, at step $s$,
\begin{equation}
\delta t(s)=\frac{1}{\nu P(s)},
\end{equation}
where the frequency $\nu$ fixes time units.

In consequence, the real-time average variation in $P(s)$ can be
obtained from
\begin{equation} \label{a1}
\frac{d\ov P }{dt}
\approx \ov {\left[ \frac{P(s+1)-P(s)}{\delta t(s)}\right] }=
\nu \ov {[1-w(s)]P(s)} =
\nu (1-\mu) \ov P ,
\end{equation}
where  overlines  indicate  average   over  realizations  of   the
stochastic  process  $w(s)$.   To  obtain  this  average evolution
equation  we  have  used  eqn  (\ref{Pest}),  and  have taken into
account  that  $w(s)$  and   $P(s)$  are  independent   stochastic
processes at each step $s$, so that $\ov {w(s)P(s)} =\ov w(s)  \ov
P(s)=\mu \ov P(s)$.

If $\nu$  is identified  with the  birth rate  per individual  and
unit  time,  the  product  $\nu\mu$  is  the  mortality rate.  For
constant $\nu$ and $\mu$, the solution to eqn (\ref{a1}) is
\begin{equation}
\ov P(t) = P_0 \exp[\nu(1-\mu) t],
\end{equation}
with $ P_0$ the initial population.

\section{Continuous approximation to the distribution of family
sizes}
\label{ap2}

Under the action of both birth and death events, the evolution  of
the average number of families of size $i$, $\ov n_i$, is given by
eqns (\ref{Simon2i})  and (\ref{Simonm}).  These equations  can be
approximately solved assuming that  the solution varies slowly  on
$s$  and  $i$,  so  that   these  two  discrete  variables   admit
replacement  by  continuous  variables  $z$ and $y$, respectively.
Accordingly, $\ov  n_i(s)$ is  replaced by  a continuous  function
$n(y,z)$.  The  approximation  is  based  on  the  expansions $\ov
n_{i\pm 1} (s) \equiv n(y,z)\pm \partial_y n(y,z)+\cdots$ and
$\ov n_i (s+1)  \equiv  n(y,z)+   \partial_z  n(y,z)+\cdots$,
to   be introduced in eqns (\ref{Simon2i}) and (\ref{Simonm}) at
different truncation orders.

\subsection{First-order approximation}

To the first order, we obtain the differential equation
\begin{equation}    \label{e1}
\frac{\partial n}{\partial z}+\frac{1-\alpha-\mu}{P_0+(1-\mu)z}
\frac{\partial}{\partial y} (yn) =0
\end{equation}
for   $y>1$.   The   contribution   for   $y=1$,   given   by  eqn
(\ref{Simon3i}), can  be incorporated  to (\ref{e1})  either as  a
boundary condition or as a singular inhomogeneity in the  equation
itself. The latter yields
\begin{equation}    \label{e2}
\frac{\partial n}{\partial z}+\frac{1-\alpha-\mu}{P_0+(1-\mu)z}
\frac{\partial}{\partial y} (yn) = \alpha \delta (y-1),
\end{equation}
where $\delta (x)$ stands for the Dirac delta distribution. In the
following, we obtain and analyse the solution to this equation.

Introducing the auxiliary variables
\begin{equation}    \label{var}
\xi = \ln \left(1+ \frac{1-\mu}{P_0} z\right), \ \ \ \ \ \ \ \
\eta=\ln y,
\end{equation}
eqn (\ref{e2}) becomes
\begin{equation}    \label{e4}
\frac{\partial f}{\partial \xi}+\frac{1-\alpha-\mu}{1-\mu}
\frac{\partial f}{\partial \eta} = \frac{\alpha}{1-\mu} \delta
(\eta).
\end{equation}
with $f(\eta,\xi) \equiv y n(y,z)$. This is a one-dimensional wave
equation,  with  ``spatial''  variable  $\eta$  and   ``temporal''
variable $\xi$.   It describes  shape-preserving advection  of the
``density''   $f$   with   constant   velocity  $v=(1-\alpha-\mu)/
(1-\mu)$, subject  to the  action of  a point  source of intensity
$\alpha/(1-\mu)$ at $\eta=0$. In our problem, thus, this  equation
makes sense  for $v>0$  ($\mu <  1-\alpha$) since  $\eta$ must  be
non-negative ($y\equiv i \geq 1$). The general solution for  $\eta
\neq 0$ is given by  an arbitrary combination of functions  of the
form $f(\eta, \xi)= f_0 (\eta-v\xi)$, where $f_0$ is in  principle
arbitrary. The combination must be  chosen in such a way  that the
boundary and initial conditions  are satisfied. In our  case, this
is achieved  with a  combination of  two such  functions.   In the
original variables, the solution is given piecewise as
\begin{equation}
n(y,z)=\alpha \frac{ P_0+(1-\mu)z}{1-\alpha-\mu}\
y^{-1-(1-\mu)/(1-\alpha-\mu)}
\end{equation}
for $y<y_T(z)$, and
\begin{equation}
n(y,z)= y_T^{-1} n(y/y_T(z),0)
\end{equation}
for $y>y_T(z)$. Here, $n(y,0)$ is the initial distribution. The
transition point between the two pieces is located at
\begin{equation}
y_T(z) =  \left(1+\frac{ 1-\mu }{P_0}z
\right)^{(1-\alpha-\mu)/(1-\mu)}.
\end{equation}
These expressions give the first-order continuous approximation
$n(y,t)$ to the distribution of families by size.

The  average  total  number   of  surnames  in  the continuous
approximation is defined in (\ref{avNc}). For the first-order
approximation we find
\begin{equation}
\ov N(z)= \int_1^\infty n(y,z) dy= \alpha \frac{
P_0+(1-\mu)z}{1-\alpha-\mu}\int_1^{y_T} y^{-\zeta} dy+
\int_{y_T}^\infty n(y/y_T,0) \frac{dy}{y_T}.
\end{equation}
The  last  integral  is  clearly  equal  to  the initial number of
surnames, $N_0$, as may be immediately  realized by the  change of
variables $y/y_T\to y$. Explicit calculation of the first integral
shows that $\ov N(z)=N_0+\alpha  z$, i.e. exactly the  same result
as for the case of $\mu=0$, eqn (\ref{Ns}). As a function of time,
we have
\begin{equation}
\ov N(t) =N_0 + \frac{\alpha P_0}{1-\mu} \{\exp[ \nu (1-\mu)
t]-1\}.
\end{equation}

\subsection{Second-order approximation}

Truncation to the second order in the continuous approximation to
eqns (\ref{Simon2i}) and  (\ref{Simonm}) yields
\begin{equation}    \label{seg}
\frac{\partial n}{\partial
z}+\frac{1-\alpha-\mu}{P_0+(1-\mu)z} \frac{\partial}{\partial y}
(yn)+\frac{1-\alpha+\mu}{2[P_0+(1-\mu)z]}
\frac{\partial^2}{\partial y^2} (yn)=\alpha \delta (y-1).
\end{equation}
Unfortunately, this   equation  cannot  be analytically solved for
arbitrary initial conditions.  However, a particular  solution can
be  found  in  the  form  of  a  separate function, $n(y,z) =
h(y)P(z)=h(y) [P_0+(1-\mu)z]$. Comparison with eqns (\ref{ni}) and
(\ref{sol1}) suggests  that  this  particular  solution  will
correspond to the long-time asymptotic evolution. It reads
\begin{equation}
n(y,z) = \frac{\alpha P(z)}{1-\alpha-\mu} \left(2
\frac{1-\alpha-\mu}{1-\alpha+\mu} \right)^{\zeta-1}
y^{-1}U\left(\zeta-1,0, 2 \frac{1-\alpha-\mu}{1-\alpha+\mu}  y
\right),
\end{equation}
where $U(a,b,x)$ is the logarithmic Kummer's function  (Abramowitz
\& Stegun, 1970).

\end{document}